%% file: template.tex
\documentclass[preprint,hideappendix,journal]{vgtc}            

\onlineid{0}

\vgtccategory{Research}

\vgtcpapertype{theory/model}

\title{VisActs: Describing Intent in Communicative Visualization}

\author{
    \authororcid{Keshav Dasu}{0000-0002-7689-4368},
    \authororcid{Yun-Hsin Kuo}{0009-0000-1891-8993}, and
    \authororcid{Kwan-Liu Ma}{0000-0001-8086-0366}
}

\authorfooter{
  \item Keshav Dasu, Yun-Hsin Kuo, and Kwan-Liu Ma are with the University of California, Davis. E-mail: 
\{kdasu, yskuo, klma\}@ucdavis.edu

}

\abstract{Data visualization can be defined as the visual communication of information. One important barometer for the success of a visualization is whether the intents of the communicator(s) are faithfully conveyed. The processes of constructing and displaying visualizations have been widely studied by our community. However, due to the lack of consistency in this literature, there is a growing acknowledgment of a need for frameworks and methodologies for classifying and formalizing the communicative component of visualization. This work focuses on intent and introduces how this concept in communicative visualization mirrors concepts in linguistics. We construct a mapping between the two spaces that enables us to leverage relevant frameworks to apply to visualization. We describe this translation as using the philosophy of language as a base for explaining communication in visualization. Furthermore, we illustrate the benefits and point out several prospective research directions.
} 


\graphicspath{{figs/}{figures/}{pictures/}{images/}{./}} 

\usepackage{tabu}                      
\usepackage{booktabs}                  
\usepackage{lipsum}                    
\usepackage{mwe}                       

\usepackage{mathptmx}                  
\usepackage{microtype}                 
\PassOptionsToPackage{warn}{textcomp}  
\usepackage{textcomp}                  
\usepackage{times}                     
\usepackage{cite}                      
\usepackage{tabu}                      
\usepackage{tabularx}

\usepackage{amsmath,amssymb}
\usepackage{color}
\usepackage{colortbl}
\usepackage[normalem]{ulem}
\usepackage{array}
\usepackage{enumitem}

\newcommand*{\tabIndent}{\hspace*{0.5cm}}%

\usepackage[font={small,sf}]{caption}
\usepackage[font={scriptsize,sf}]{caption}

\begin{document}


\firstsection{Introduction}

\maketitle

\input{sections/01_introduction}
\input{sections/02_vislanguage}

\input{sections/03_visintents}

\input{sections/03_02_SAT_Background}

\input{sections/04_00_visacts}

\input{sections/04_01_visactsapplied}
\input{sections/05_discussion}

\input{sections/06_conclusion}

\bibliographystyle{abbrv-doi-hyperref}

\bibliography{template}

\appendix 

\section{About Appendices}
Refer to \cref{sec:appendices_inst} for instructions regarding appendices.

\section{Troubleshooting}
\label{appendix:troubleshooting}

\subsection{ifpdf error}

If you receive compilation errors along the lines of \texttt{Package ifpdf Error: Name clash, \textbackslash ifpdf is already defined} then please add a new line \verb|\let\ifpdf\relax| right after the \verb|\documentclass[journal]{vgtc}| call.
Note that your error is due to packages you use that define \verb|\ifpdf| which is obsolete (the result is that \verb|\ifpdf| is defined twice); these packages should be changed to use \verb|ifpdf| package instead.

\subsection{\texttt{pdfendlink} error}

Occasionally (for some \LaTeX\ distributions) this hyper-linked bib\TeX\ style may lead to \textbf{compilation errors} (\texttt{pdfendlink ended up in different nesting level ...}) if a reference entry is broken across two pages (due to a bug in \verb|hyperref|).
In this case, make sure you have the latest version of the \verb|hyperref| package (i.e.\ update your \LaTeX\ installation/packages) or, alternatively, revert back to \verb|\bibliographystyle{abbrv-doi}| (at the expense of removing hyperlinks from the bibliography) and try \verb|\bibliographystyle{abbrv-doi-hyperref}| again after some more editing.

\end{document}

%% file: sections/01_introduction.tex
Data visualization is a vast and growing field. In this paper, we focus on the subspace of communicative visualization, a space that is concerned with the explanatory side of data visualization and is often what the average person is exposed to.
In this space, the communicative goals can range depending on the designer's intentions and audience~\cite{lee2012beyond,ojo2018patterns,Bako2022}. 
The role data visualizations play in communication varies where some designers use them to supplement their written and spoken messages, whereas others recognize them as an entirely effective mode of conveying the message~\cite{segel2010narrative,stolper2016emerging,quadri2022}. 
Consequently, we are seeing wide usage of data visualization in industry and academia to communicate increasingly diverse and sophisticated messages.
The complexity and diversity of interaction data visualization usage suggest that we could benefit from looking at it as a rich language.
Developing frameworks from this perspective could then allow us to glean insights from naturally occurring experiments in practice and enable research that can guide future practice. 
Given the growing sophistication of visual communication, there is value in exploring the relevance to data visualization of frameworks developed by linguists and language philosophers.
Some initial frameworks to examine are speech act and discourse theory; speech act  because it distinguishes between the what is 'said', the intentions of the speaker, and the consequences of what is said in how it is processed by listening.  Discourse theory because it examines how our communication is shaped by external structures.

Here the focus is on intent. Designers are tasked with creating and evaluating visualizations for targeted audiences.
These audiences  have varying motivations to engage with the presentation and with differing levels of prior knowledge of the subject matter.
Designers have a wide array of intents.
These intents range from journalists attempting to {\it inform} readers, teachers trying to {\it explain} concepts, scientists attempting to {\it discover} relationships among variables, policymakers hoping to {\it persuade} the public about the rationale for a decision, a blogger seeking to {\it evoke} strong emotions and activists hoping to get volunteers to {\it act}.
How can we classify these intents in a manner that advances our ability to visualize data? 
Classifying intent is a prerequisite for determining if a visualization adequately satisfies the communicative intent of designers. 
Thus, to build and evaluate communicative visualizations we need a refined and principled language for describing communicative intent. 

Recent work in communicative visualization by Adar and Lee~\cite{adar2020communicative} tackles the question of “how do we formally describe communicative intent in visualizations?”  Their work offers an initial taxonomy for intents and enables an additional discussion on how to communicate information using visualization.
Others have also identified the importance of intent.
For example, Schoenlein et al.~\cite{schoenlein2022unifying} note a central problem in visual communication is understanding how people infer meaning from visual features.
All this points to a need to assess and understand if our communicative goals as designers are being correctly imprinted in our visual features as intended.
We posit that when considering the question of “how can we formalize intents” with regards to visualization, we can draw from the philosophy of language, particularly speech act theory~\cite{austin1975things,searle1968austin,searle1969speech,grice1957meaning,stalnaker1999context,murray2018force}. 

Our work aims to link the field of visualization to the field of linguistics and demonstrates how doing so offers a broader perspective of the space and introduces new opportunities for research that can facilitate design.
We illustrate the connection between these spaces by explaining the link between a sub-space of visualization, i.e.,  communicative visualization, to a sub-space of linguistics, i.e.,  speech act theory. We show how this relationship can help grow our understanding of communicative visualization. The insights and formalization developed there can guide us in developing a formal language for intent in visualization.
With VisActs, we offer a framework to assist in enhancing the overall communicative precision of a data-driven visualization. 
Our framework complements task-based design analysis by examining the design at a granular level, providing an approach to understanding how designer intent affects low-level design decisions.

In this paper, we (a) propose VisActs, which leverages speech act theory, as a framework for studying intent in visualization and (b) delve deeper into intents by (i) identifying a set of oft-encountered vis design intents (ii) illustrating the relationship between the intent and visualization (examples of same content visualized differently based on the intent), (iii) showing how the mode of achievement creates a mesh of intents, (iv) showing the impact of context and conventions on how intent is realized (or made difficult to achieve).

%% file: sections/02_vislanguage.tex
\section{Data Visualization as a Language}
There is an ongoing discussion on design as communication~\cite{crilly2008design,krippendorff1989essential,barnard2013graphic,george2002analysis}
and there is a body of  work~\cite{crilly2008design,barnard2013graphic,george2002analysis} that gives credence to viewing visual design as communication.
In this work, we engage with this ongoing discussion and identify the implications and research directions that emerge from viewing visualization design as a language.

Visualizations share many commonalities with language as they both express what we observe to others. 
The goal of visualization, like ordinary speech, often goes beyond presenting facts to achieving actions or outcomes. In any case, how data is visualized can alter how the original context is perceived, e.g., visualizing the uncertainty in data~\cite{greis2018uncertainty,hullman2019authors}.

Treating visualization as a language has been considered, although exploring the value of this association and what it affords is limited.
Purchase et al.~\cite{purchase2008theoretical}  have explicitly made these connections, and briefly describe the use of linguistic theory, namely pragmatics, to provide an over-arching framework for information visualization.
They comment on the relationship between visualization and language and discuss how information visualization should rely on a multitude of theories rather than a singular theory.
Hullman and Diakopoulos~\cite{hullman2011visualization} study the application of linguistic-based rhetoric in narrative visualizations.
Several others have presented theoretical visualization frameworks~\cite{liu2008distributed,chen2010information,kindlmann2014algebraic,parsons2018conceptual} and  implicitly imply that visualization is a language.
They elegantly demonstrate how applying frameworks from spaces such as distributed cognition, information theory, an algebraic basis, or conceptual metaphor theory can contribute to the betterment and improved understanding of the use of visualization.
 
A vocabulary is the body of words used in a language.
If we are to claim visualization is a language, then its vocabulary would be visual encodings.
This association has been observed by Wilkinson~\cite{wilkinson2012grammar}, who identified general rules that govern the creation and presentation of data graphics and presented a structure within which these rules might be operationalized efficiently. 
He supposed that if the grammar is successful, it should be possible to reduce any data visualization problem into a graphic utilizing the rules outlined. 
Grammar-based visual encodings, as well as declarative languages~\cite{mackinlay1986automating,hanrahan2006vizql,heer2010declarative,bostock2011d3,li2019p5,satyanarayan2016vega}, arose out of a need to fluidly and precisely articulate a set of intents for communicating data.
These works provide formalized approaches for describing tables, charts, graphs, maps, and tables and give credence to treating visualization as a language.

In summary, researchers have recognized that visualization is a language and that it would benefit from formalizing the relationships to languages. If we are to treat the totality of visualization as a language and apply linguistic frameworks, we would have common ground for discussion and understanding of the properties of expressing visualizations, thereby facilitating the development of the field.

 We present an approach for translating relevant theoretical frameworks from the space of linguistics into visualization. We develop a mapping between a subspace of visualization and linguistics \textit{to illustrate} the potential for more work in this direction and immediate benefits. Our focus is on the intent of the designer. We propose a theoretical structure for both describing and assessing the intents of visualization designers and the respective consequences.

The motives of the designer of a visualization -- to achieve actions or outcomes -- and the impact of the visualization on perceptions, whether intended or not, should be considered while developing theoretical frameworks for studying visualizations.
A framework to capture how we design interactive visualizations and their effects can be obtained by developments in speech act theory.
Speech act theory, a sub-field of pragmatics and linguistics, studies how words are used to both present information as well as carry out actions.
We describe a mapping of speech act theory into visualization and offer a theory on visualization acts, \textit{VisActs}.
This framework will be linguistically motivated using the foundation of speech act theory but made relevant for the space of visualization.
That is, it must account for fundamentally how visual elements are processed and interpreted, which delves into semiotic theory.
Furthermore, it must take into account both the conventional content and the underlying principles of data visualizations. 
Finally, such a theory should also offer testable predictions about the kinds of \textit{VisActs} performed across the space of visualization.
Particularly, it should offer the ability to assess how our intents manifest within visualization and their respective consequences.

Next, we delve into intent in visualization. Subsequently, we explain speech act theory and how it relates to communicative visualization. 
This is immediately followed by our introduction of \textit{VisActs}, a translation of speech act theory contextualized for visualization researchers.
We ground the relevance and application of this translation through a series of examples, followed by a discussion about our mapping.

%% file: sections/03_visintents.tex
\begin{table*}[t]
\small
\centering
\begin{tabular}{p{0.01\textwidth}>{\raggedright}p{0.2\textwidth}>{\raggedright}p{0.35\textwidth}p{0.35\textwidth}}
    \toprule
    \multicolumn{2}{p{1.5625in}}{\textbf{Speech Act Theory Taxonomy}} & 
    \textbf{Description} &
    \textbf{Translation into Visualization}  \\
    \arrayrulecolor{black!30}\midrule
    
    \multicolumn{2}{p{1.5625in}}{\textbf{Fundamental Concepts}} & 
    &
    \textbf{A theoretical framework for describing visualization designer's intents.}   \\
    
    &
    Locutionary Act & 
    The utterance of a phrase. \textit{What is heard.} &
    To show data. \textit{What is shared.} (Section 5.2) \\
    
    &
    \tabIndent Phatic Act &
    An utterance of words which has meaning &
    The selection of data. "Data Act"\\
        
    &
    \tabIndent Propositional Act &
    The act of expressing the proposition, the content &
    Expression of data via analysis."Analytic Act" \\
    

    &
    \tabIndent Sentence Type &
    The type of sentence (e.g. declarative, exclamatory, etc.) has an impact on the force of an utterance. &
    The visualization type (i.e. informative, instructive, narrative, explorative, and subjective) has an effect. \\
    
    &
    Illocutionary Act &
    The utterance of a phrase with an intention. \textit{What is intended.} &
    The design of visualization with an intention. \textit{What is seen.} (Section 5.3) \\
    
    &
    Perlocutionary Act &
    Effect utterance had on the listener. \textit{The consequence.} &
    The effect a visualization has on the viewer.\textit{What is understood.} (Section 5.4) \\
    
    &
    Context &
    A cluster of actual states of affairs or various events related to the utterance. & 
    The objects or entities which surround a focal event and provide resources for its appropriate interpretation.\\
    
    &
    Convention &
    Societal rules and norms govern countless behaviors. &
    Visualization design abides by these as well. \\
    
    \midrule
    
    \multicolumn{2}{p{1.5625in}}{\textbf{Illocutionary Force}} & 
    The speaker's intention behind the utterance &
    The designer's design rationale behind their visualization. (Section 5.4)\\
  
     &
    Illocutionary Point (IP) & 
    The point or purpose of a type of illocution  & \\
    
     & 
    \tabIndent Assertive Point &
    Convey information. The utterance that informs how things are. Either true or false. &
    To visually state, claim, or suggest something is the case. \\
    
    & 
    \tabIndent Commissive Point &
    Make a commitment. &
    The guarantees of what a visualization will offer and abide by (data authenticity). \\
    
    & 
    \tabIndent Directive Point &
    Attempts by the speaker to get the hearer to do something. &
    Engaging or motivating the viewer to do something via the visualization. \\
    
    & 
    \tabIndent Declarative Point &
    Create a new state. Utterances that change the world by representing it as being changed. &
    Data transitions or transformation as well as predictive visualizations. \\
    
    & 
    \tabIndent Expressive Point &
    Reveal the speaker's attitude or emotion towards a particular proposition. &
    Revealing personal bias or sharing personal opinions through visualization. \\
    
    &
    Degree of strength of IP &
    These points can be achieved with different degrees of strength. &
    Degree of the design's effort to convey an IP through the visualization.\\
    
    &
    Mode of achievement of IP & 
    The various means a speaker utilizes to achieve the IP of an utterance. &
    The means a designer employs to communicate the IP of the visualization.\\
    
    &
    Propositional Content Conditions & 
    A limitation on the nature of the state of affairs for an IP. &
    Each IP has conditions that need to be met for the illocution to register. \\
    
    &
    Preparatory Conditions &
    A state of affairs that is presupposed is a necessary condition for the non-defective employment of the force. &
    Assumptions the designer makes about a viewer when employing a particular force.  \\
    
    &
    Sincerity Conditions & 
    The psychological state of the speaker concerning the IP. &
    The designer and the viewer take the visualization and all its content as intentional. \\
    
    &
    Degree of strength of sincerity conditions & 
    The strength of the psychological state the speaker commits to when employing an IP. &
    \\
  
   \midrule

\end{tabular}
\caption{Key Information on speech act theory concepts, applications, and meaning in visualization.}
\label{tab:speech_vis}
\end{table*}

\section{Communicative Intent in Visualization}

Communicative visualizations are created for a broad audience and represent the majority of visualizations that the public encounters. As stated earlier, communicative visualization occurs in a range of settings including journalism, education, museums, and public policy discussions. The audience differs in terms of their backgrounds, familiarity with the subject, and initial level of interest in the topic. This is in sharp contrast to visualizations that are designed for analysts or domain experts, where the designer has an understanding of their audience’s prior knowledge and has some assurance that the expert will use or attempt to use the visualization tools. Furthermore, the designer’s intent is to provide visualizations that facilitate a set of specific tasks described by the experts.   The diversity in the audience for communicative visualization makes it important to understand intent.    

To start, we can consider intent as what the designer, or one who puts forth information, would like to be conveyed and communicated.  
The intent here would closely parallel the intent of a speaker in routine life.
For example, if a child asks her mother while eating soup ``is there any salt?'', her intent could be to request some salt. 
However, what she stated may also be a query about the availability of salt. 
As this example illustrates, the intent of the speaker may not be perceived by the recipient and the desired outcome fails to occur.

\begin{figure}[t]
	\includegraphics[width=\columnwidth]{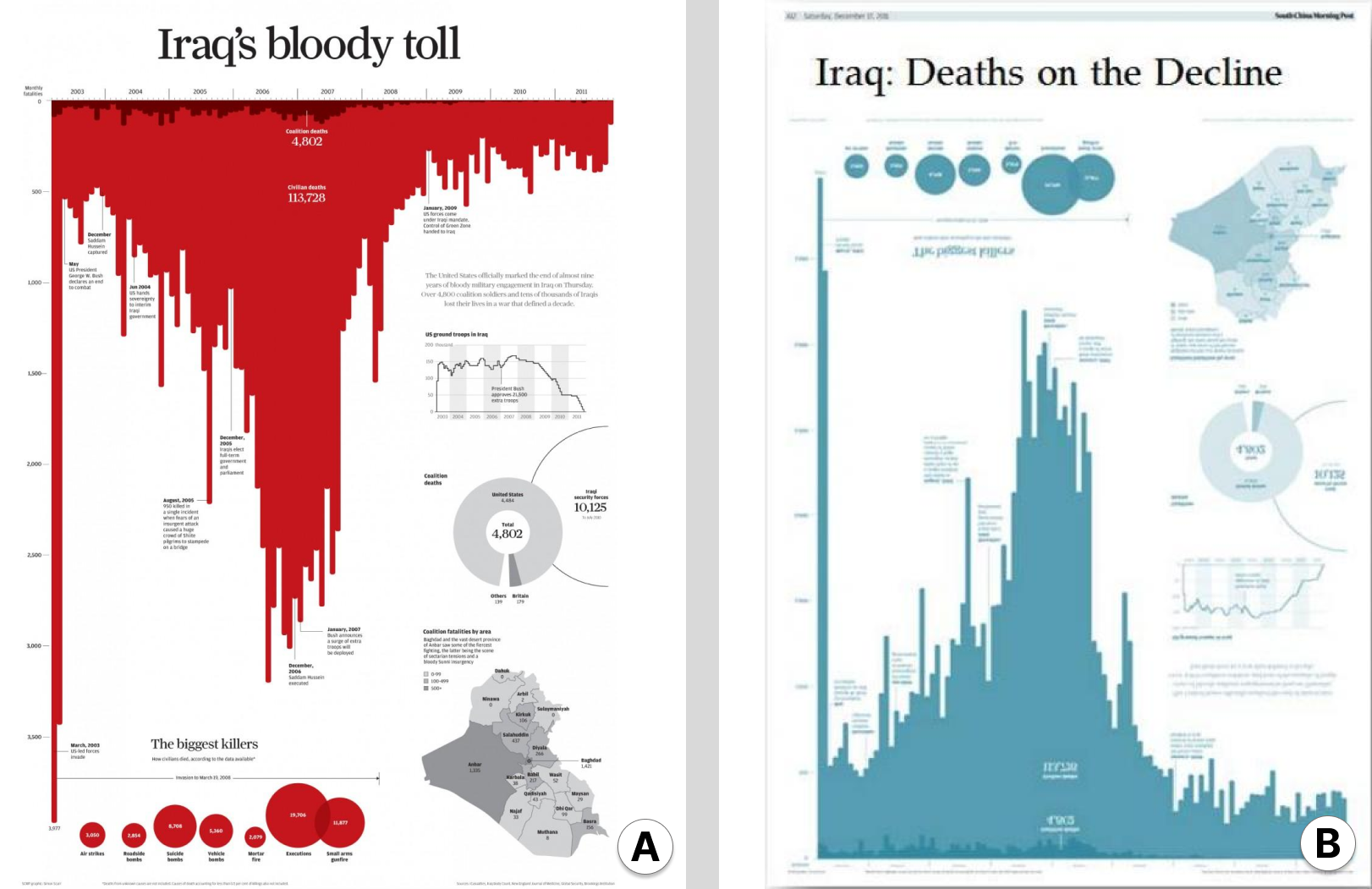}
  	\caption{ (a) depicts a visualization, by Simon Scarr~\cite{IraqsBloodyToll}, with a possible intention to terrorize the audience how devastating the Iraq War was. (b) is a revision of the same visualization by Andy Cotgreave~\cite{AndyCotgreave2023}. He adjusts the design so it is potentially received as more ``neutral.''}
  	\vspace{-3mm}
	\label{fig:designer-intent}
\end{figure}

Similarly, let us consider a visualization designer who creates a chart about war.
One intent of such a visualization could be to terrorize~\cite{ojo2018patterns} the audience to act in protest of the war, for example, in~\autoref{fig:designer-intent}-(a) the designer could produce a visualization that appears dramatic and possibly violent.
However, the same data with another intent can be visualized differently.
As seen in ~\autoref{fig:designer-intent}-(b), where a more ``neutral'' design with a possible intent to inform the public about the war.


In both situations, we want the communicative intentions to be received accurately. However, the diversity of the audience may make the intent of the speaker or the designer different from what is perceived by the receiver.  The similarity of the communication challenges in these domains stems from the nature of the audience. Hence, it seems fruitful to leverage the findings in speech act theory to inform design practices in communicative visualization.
In the space of visualizations, intent has not been formally defined.  In the following subsections, we will offer some formalism.  Fundamentally, it is useful to consider intent from two perspectives, user-intents and designer-intents.

\subsection{User Intent}
Dimara and Perin~\cite{dimara2019interaction}, when characterizing interaction in visualization, offer a definition of the term \textit{data-oriented intent}.
In their work, they describe interaction as a goal-oriented activity with a data-oriented intent.
They survey the literature that describes intents from the perspective of a user~\cite{abras2004user,gotz2009characterizing,lee2012beyond,liang2010exploratory,liu2010mental,quadri2022}.
They find that the visualization literature classifies \textit{intent}, from the perspective of a user, as a goal, task, or problem. 
User intent could be to explore high-level data, gain insights, or gain multiple perspectives of data. 
The intent of a user can also be to collect and correct data, identify insights, or make decisions.  
In this literature, the intent has been described and identified at a low operation level (e.g., altering representations and collecting data) as well as at a higher level (e.g., information foraging, sense-making, and knowledge creation).

 As designers, we tend to remove ourselves and our own intentions and in a manner treat ourselves as an outside entity constructing an interface from the perspective of a user to satisfy the user's intentions. 
Our research field has identified a variety of ways to adapt~\cite{2021_ivi_intent} our visualizations and systems to the intentions of the user.
Designers spend a lot of time describing user intentions in terms of workflows and strategies and curating systems accordingly.
A goal as a designer is to create interfaces that enable users to effortlessly express their intentions through the data.

\subsection{Designer Intent}
In the spaces of narrative visualization and data storytelling, there are many papers~\cite{stolper2016emerging,segel2010narrative,ma2013engaging,boy2015storytelling,lee2015more,dasu2020sea} that provide frameworks and methodologies for communicating narratives.  Although these papers do not explicitly define or identify the designer's intent, they subsume a diffuse concept of intent.

Bako et al.~\cite{Bako2022} assessed how designers utilize and apply data visualization examples by determining example usefulness, curation practices, and design fixation.
Their work gives us methods for capturing the designer's intent as they begin the process of developing their visualizations.
Often designers may not be able to articulate the form of what they intend to communicate.
Examples are an effective way to express and collage together this form.
Another explicitly identified type of designer intent is artistic intent. 
Artistic intent often disregards functionality, making some works unintentionally incomprehensible. 
Lau and Moore~\cite{lau2007towards} offer a conceptual model for incorporating these intentions formally.
Intent may also have social motivations such as coordination, collaboration, or presentation to an audience. 

{\color{blue}}

Recently, Adar and Lee~\cite{adar2020communicative,Lee2023} put forth a definition and a taxonomy for expressing intents in communicative visualization.
To our best knowledge, their work is the only attempt to provide a formal classification of \textbf{intents} that are broadly applicable.  They proposed a cognitive taxonomy in which they frame intents to be of the structure, “The viewer will [verb] [noun].” 
Verbs are selected from a specified set of cognitive constructs; nouns are from a set of specified knowledge dimensions. 
Their primary claim is that a good language for describing intents is the language of learning objectives.
They assert that the advantages of using learning objectives are: ``(1) being capable of describing objectives regardless of what the viewer wants; (2) allowing for a designer to create specific tests to validate if a visualization leads to a viewer achieving an objective; (3) finding out if the objectives are achieved both when a viewer is looking at a visualization and when it is taken away''~\cite{adar2020communicative}.
A limitation of their work is that it restricts the intent of the designer to educate the audience. On the other hand, this is the only paper that provides some formalization of the designer's intent.

We seek to add to the discussion of designer intent by providing an alternative perspective for viewing intent in visualization and demonstrating how this perspective can assess and analyze designer intent at a granular level.

\subsection{Challenges with Intentions}
Our intentions can manifest in many forms in data visualization, especially as our communicative goals evolve and become more nuanced. 
Through examination of research~\cite{adar2020communicative,lee2022affective,ojo2018patterns,schoenlein2022unifying} that addresses the various types of intentions in communicative visualization, we highlight the following set of intentions to illustrate these forms; however, similar to spoken word, they are not limited to this set.

\begin{enumerate}[noitemsep,topsep=2pt,parsep=0pt,partopsep=0pt]
    \item \textbf{Inform}: the intention is to have the audience be \textit{aware} of a concept.
    \item \textbf{Educate}: the intention is to have the audience \textit{understand} a concept.
    \item \textbf{Emote}:  the intention is to \textit{illicit} an emotional response from the audience. (enjoy, anger, sadness, etc.) from the presentation of the concept.
    \item \textbf{Provoke}: the intention is to get the audience to \textit{react} or\textit{ take action} to the concept presented.
    \item \textbf{Discovery}: you are \textit{obscuring} information on purpose so that people work for it and through that work, they \textit{gain} some insight that can only be gained through this process.
\end{enumerate}

It is known to be challenging, with absolute certainty, to derive an individual's original intentions behind an action~\cite{austin1975things} and likewise a data visualization. 
However, through other contexts, structures, and cues, it is possible to infer a close approximation of their intent.
Linguistics has spent time studying both pragmatic and semantic structures in language as a means to accurately gauge intent, which has attracted interest from law practitioners as well as the NLP (Natural Language Processing) community.  We hope frameworks for analyzing data visualizations, such as the one proposed here, can help the rapidly developing communicative visualization subspace and its relationship with ML4VIS.

%% file: sections/03_02_SAT_Background.tex
\section{Speech Act Theory Fundamentals and Terms}
In this section, we will review our translation process and provide additional information on the terminology.
\autoref{tab:speech_vis} contains the terminologies that we translate and contextualize for data visualization. 

The field of speech act theory examines how words are not only used to convey information but also to carry out actions.
Many philosophers and linguists study speech act theory to gain insights and a better understanding of how we communicate.
A speech act can be described as something that is expressed by an individual that not only offers some information but also performs an action.

The initial foundation of speech act theory was introduced by J.L Austin~\cite{austin1975things} and the theory has since been developed and expanded by several other scholars~\cite{grice1957meaning,searle1969speech,murray2018force}.
Austin introduced the terms \textit{locutionary}, \textit{illocutionary}, and \textit{perlocutionary} acts. 
Where locutionary act is the utterance of the phrase, illocutionary is what was meant or intended by the utterance, and perlocutionary act is the effect the utterance has upon the listener. 
These terms of locutionary, illocutionary, and perlocutionary can, respectively, be thought of as: what is being put forth, how is it being put forth, and what does putting it forth achieve?  

\subsection{Forces}
Classical speech act theory~\cite{austin1975things,searle1969speech,searle1985expression,grice1957meaning} introduces the idea that our utterances, words with some meaning that we put forth, contain a variety of forces.
Grice~\cite{grice1957meaning} introduced the concept of speaker meaning, a speaker attempts to get the audience to believe something by relying on the audience to take the intention of the speaker as a reason for belief.
Grice finds that in order for a speaker's meaning to occur, the speaker must first intend to produce an effect on an audience and also intend that this very intention be recognized by that audience.
Next, the speaker must also intend this effect on the audience to be produced at least in part by their recognition of the speaker’s intention. 
Speech act theory recognizes~\cite{cohen1964illocutionary,searle1985speech,murray2018force} that an illocutionary force contains the intent of the speaker.
Namely, illocutionary force is the intended message a speaker assigns to a sentence they utter.
Searle and Vanderveken~\cite{searle1985speech} assert that the force in speech is comprised of 7 parts; illocutionary point (IP), degree of strength of the IP, mode of achievement, propositional content conditions, preparatory conditions, sincerity conditions (SC), strength of SC. 
The illocutionary point can be of the following forms: assertive, commissive, directive, declarative, and expressive.

Neo-Gricean theories modify Grice’s principles to some extent.
From these modifications, we are given relevance theories~\cite{sperber1986relevance} as well as modifications to forces allowing for more focus on the speaker's intention.
In this work, we use the Neo-Gricean analysis~\cite{bach1979linguistic,cohen1979elements} as a basis for our mapping between communication in visualization and speech act theory.
Mapping these forces into visualization requires careful consideration of what is consistent and what additionally needs to be factored in.
Murray and Starr~\cite{murray2018force} propose that the force of an utterance is its communicative function.
They examined Grice's definition of communicative intention~\cite{grice1957meaning} and found that it did not consider how signals are used to coordinate communications. Although Murray and Starr~\cite{murray2018force} state that the approach we adopt does not address how agents use signals to coordinate, in the context of visualization, we fill this gap using semiotic theory.
As visualization designers, we make use of visual signals which is explained by semiotic theory.
An important takeaway of semiotics~\cite{peirce1991peirce,johansen2002signs,curtin2009semiotics} is how social conventions influence meaning. 
In other words, the force of an utterance is contextual and subject to conventions.

\begin{figure}[h]
	\includegraphics[width=\columnwidth]{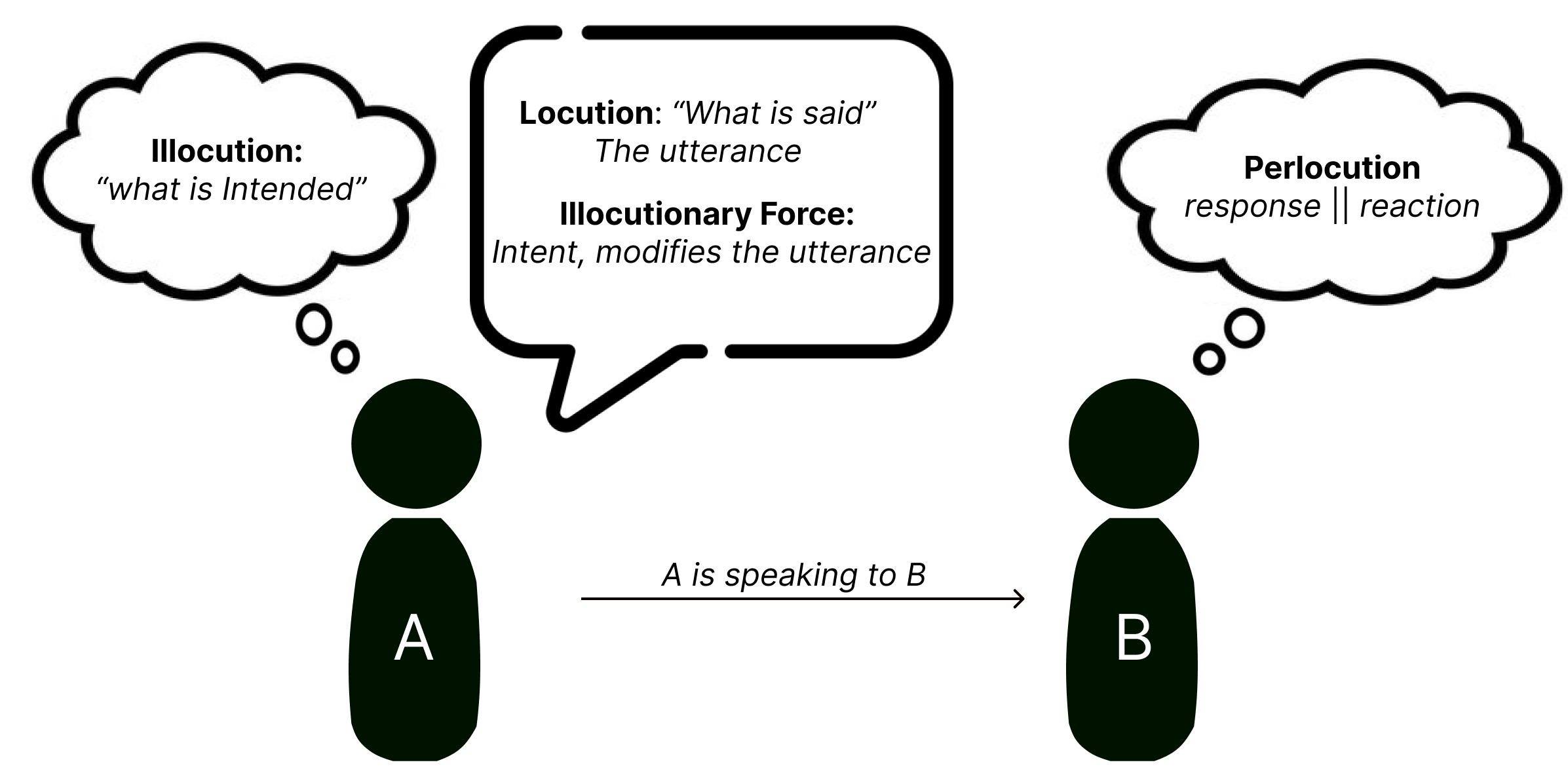}
  	\caption{Illustration of a speech act. Person A utters a phrase, a locution, to person B. This locution is what person B will hear. Person A simultaneously also performs an illocutionary act by applying a force with their intention. How person B responds or what they do after processing what Person A has said would be classified as a perlocution.}
  	\vspace{-3mm}
	\label{fig:speechact}
\end{figure}

\subsection{Speech Act Example: Alice \& Bob}
To provide a clear example of what is a speech act and what can one do with it let us observe a conversation between Alice and Bob~Fig.\ref{fig:speechact}.
\begin{enumerate}[itemindent=2.55em]
    \item[Alice:] \textit{``Would it be too much trouble for me to ask you to hand me the salt?''}
\end{enumerate}

\noindent Alice utters a sentence to Bob that is asking two questions concurrently.
The first question is if Bob is capable of passing the salt, while the second is an actual request for the salt.
The \textit{locutionary act} in this case is what was said, the literal sentence.
The \textit{illocutionary act} is what Alice means by uttering the sentence, and the \textit{illocutionary force} is the intent.
Specifically, the intention of Alice was a request for Bob to give her salt, she then issued an utterance with the illocutionary force of command to Bob.
If Bob, successfully processes this utterance and its force and proceeds to acquire and hand Alice the salt, then he has performed the \textit{perlocutionary act}.
In order for Bob to identify the illocutionary force and corresponding act, Bob must either passively or actively factor in relevant contextual information or social conventions and practices to help determine what Alice's intents are.

%% file: sections/04_00_visacts.tex
\section{VisActs}
The speech act was developed for understanding and examining how we communicate, specifically with words and speech; however, we find it can be extended past words for visual communication.
Visualization is becoming a dominant mode of communication. 
As visualization becomes more complex for expressing and communicating data, it will inherit the challenges of languages. 
At one level, it has the ability to express more precisely but concurrently opens itself up to more ambiguity and multiple ways to be interpreted, which can have a variety of implications.

\begin{figure*}
    \centering
    \includegraphics[width= \linewidth]{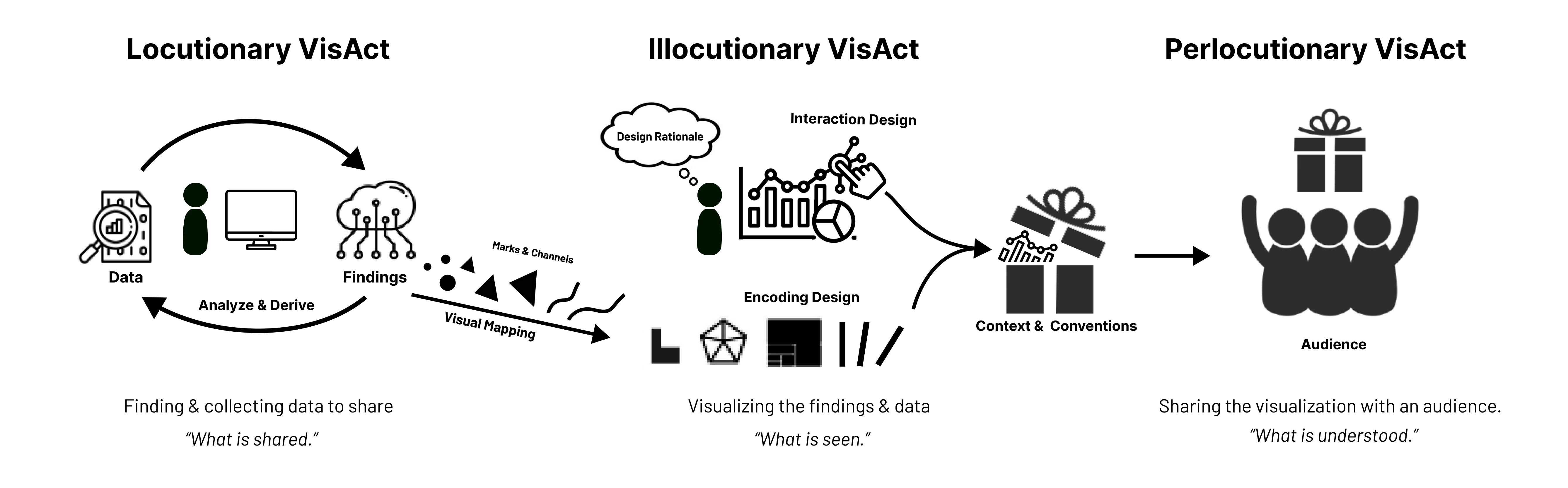}
    \caption{VisActs consists of three core actions; Locutionary, Illocutionary, and Perlocutionary acts. Through each of these actions, the designer imbues their intent into the visualization in the hope of the audience understanding and interpreting the message as intended.
    }
    \label{fig:VisActs-Process}
\end{figure*}

\subsection{Proposed Framework}
With VisActs, we borrow some concepts from linguistics to use as a foundation and then proceed to contextualize and translate how these structures apply in data visualization, specifically the communicative side of data visualization. 
In speech act theory, we can use three structures to frame intents and their effects; locutionary, illocutionary, and perlocutionary speech acts.
With this framework, we offer varying levels of depth for the analysis of designer intent, as seen in, ~\autoref{tab:speech_vis}.
Furthermore, we focus on a retrospective analysis of designer intentions in visualizations and particularly data stories to illustrate VisActs.

To begin our translation, we first contextualize each of these for visualization.
A locutionary VisAct is the \textit{data} or finding we present to the targeted user or audience.
An illocutionary VisAct is the \textit{visual representation} or \textit{encoding} this finding or takes assumes when presented to the target user or audience.
The illocutionary force or VisForce, is the \textit{design rationale} for the representation or encoding.
Lastly, the perlocutionary VisAct represents the \textit{evaluation} of the encoding design after the audience has viewed and processed it.
Through the perlocutionary VisAct, the designer gains an understanding of if their intended outcomes were met, that is the audience decoded the encodings and understood the findings or data presented as intended by the designer. 

With this framing, we have separated stages of visualization design into several bins.
In the first bin, locutionary VisAct, we focus on isolating what is the specific or derived data to convey to the audience.
This bin is not concerned with \textit{how} this data is visually represented or modified but focused on the semantic \textit{what} part of the data is being shared.
It is in the illocutionary VisAct and its accompanying VisForce, that we can begin teasing and understanding how the design impacts the communication of the data.
There are several means through which a designer can transfrom the visualization to reflect their intentions.
The two categories this work will focus on are encoding and interaction design.
However, how we communicate, design, and interpret data-driven visual content is also affected by societal conventions and other contextual information.
The goal of VisActs is to provide an alternative means to assess here how are intentions shape \textit{visually} the data we are communicating as well as better infer a designer's original intentions for producing a visualization.

\subsection{Locutionary VisAct}
For the purpose of this work, we are only concerned with data that has been identified to be shared.
VisActs does not consider data whose content is largely unknown and expected to be explored.
The Locutionary VisAct made by the designer is the process of selecting data, tasks, and initial analysis methods (i.e., data cleaning).
As these choices reflect part of the designer's intentions. 
For example, in data storytelling, this is the process of identifying the ``story pieces'' to be communicated~\cite{lee2015more}.
The data selection and modification affect the visualization design, as it may constrain what visualization options if any~\cite{quadri2022,franconeri2021science,munzner2014visualization}.
For example, hierarchy suggests depth, temporality may imply change, and spatiality could imply closeness or bonds.
Thus, we may be visualizing data as a treemap, flows, or possibly on a map.
By taking data types into account, such as nominal, categorical, numerical, and their pairings, we can begin to define a space of what representations are available to fulfill our communicative goals.

\subsection{Illocutionary VisAct}
The illocutionary VisAct is the process of designing a visualization from the data to then be shared with an audience.
This visualization may be interactive but must be data-driven.
Similar to speech acts, we are not concerned with visualizations that have no data or ``'meaning'' associated with them.
The design of both interactive and static data visualization is heavily influenced by the designer and their choices.
In this VisAct the relationship between the designer's intent, their rationale, and the resulting visualization is captured.
How the designer intends to communicate this data (e.g., to educate or persuade) may influence their design rationale.

\begin{figure*}
    \centering
    \includegraphics[width=\linewidth]{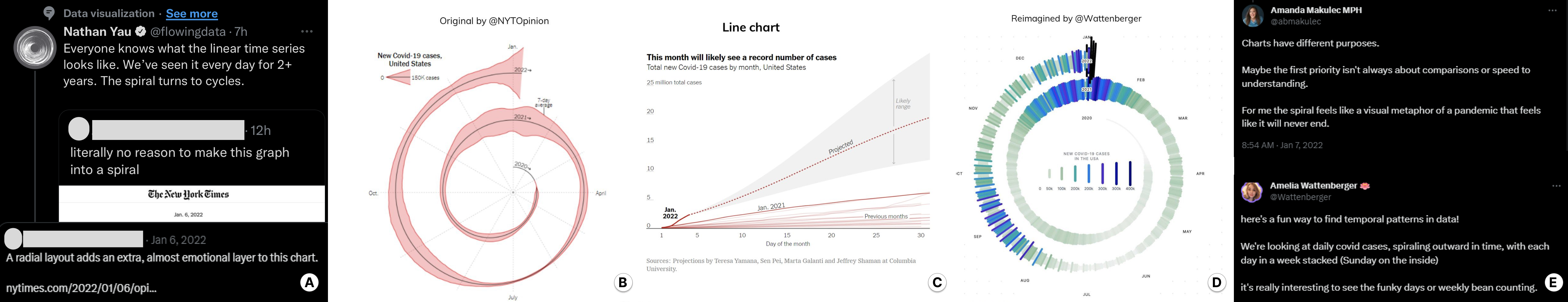}
\caption{ This figure illustrates a disconnect between an author's intention and the audience's reception. Jeffrey Shaman wrote an article~\cite{ShamanOmicron} predicting when the Omicron variant of Covid would peak with an accompanying visualization. The visualization sparked an online Twitter debate (a), with some finding the original design (b) ineffective compared to (c) a line chart. Amelia Wattenberger (d) re-imagined the design with a different intent, which garnered a more positive response, and Amanda Makulec (Executive Directory of DataVizSociety) (e) wrote a thread\cite{abmakulec} on how visualization design can address different needs or intents.
    }
    \label{fig:misaligned-intent}
\end{figure*}

The intention, or intended purpose of a design choice, is captured by an illocutionary point (IP).
VisAct IPs fall under a set of five types; assertive, commissive, directive, declarative, and expressive.
An assertive point is a visual element that either states, claims, or suggests the represented data has validity and is true.
For example, in ~\autoref{fig:misaligned-intent}c the solid red lines are making an assertive point visually \textit{stating} the current trend of Covid-19.
A commissive point sets up a guarantee between the design and the audience that it will offer either an explanation, understanding, or action.
A simple example would be a slider that sets the time window for a line chart, the guarantee being that the chart should update to reflect the selected window.
A directive point would be design choices that attempt to engage or motivate the audience to act.
Declarative points are design elements that transition the visualization into a new state or show predictions of what could transpire (i.e., animated transitions, filtering, or drill-down).
Lastly, an expressive point captures the designer's personal opinions as they appear in the visualization.
In~\autoref{fig:designer-intent}a each choice to make the chart look like blood dripping could be an expressive point; the color, inverting the y-axis, and the title of the poster.

Whereas the design rationale is referred to as the VisForce, a force that guides and nudges design decisions into what is finally seen by the audience.
A VisForce's influence appears in (1) the encoding design and (2) the interaction design.

\noindent\textbf{Encoding Design.} How we design visualizations, in terms of binding the data to the visual elements, greatly impacts how the data are perceived and understood by the audience~\cite{padilla2018decision,franconeri2021science,munzner2014visualization,Bertin2011}. 
Certain visualization design choices may elicit emotional responses from the audience, which can also help better communicate the designer's intentions to the audience~\cite{NegativeEmo2022}.

\noindent\textbf{Interaction Design.} Interaction design as it pertains to data visualization is a heavily documented~\cite{dimara2019interaction,hoque2017applying,lee2012beyond,liang2010exploratory,liu2010mental}. From these works we can surmise the following: (1) interaction design effects visually what is seen by the audience, (2) interaction design influences how the audience perceives the data, (3) and interaction design impacts audience engagement with the data.
The designer's choice of which interactions are available to the audience can steer the audience towards their goal.



\subsection{Perlocutionary VisAct}
A perlocutionary VisAct is performed by the audience.
This action informs the designer whether or not their desired outcome has transpired.
If the outcome was to \textit{call to act} over climate change by signing a linked petition and they received a signature that would be a success.
However, if the outcome was to \textit{educate} museum visitors on metagenomics~\cite{dasu2020sea} via an interactive system and the majority of visitors failed to understand the system, then it was unsuccessful.
The granularity of success and failure, much like with evaluation is up to the designer to classify.
In the context of data visualization and research, this stage is evaluating what was understood but the audience and how that aligns with what the designer intended to transpire~\cite{Prateek2023,franconeri2021science,munzner2014visualization}.

\subsection{Convention}
Social conventions are rules and norms that govern countless behaviors we all engage with every day.
Bicchieri~\cite{bicchieri2005grammar} defines conventions as descriptive norms that have endured the test of time. 
She states that if one’s main objective is to coordinate with others, and the right mutual expectations are present, people will follow whatever convention is in place.
In visualization design adhering to and following conventions is necessary for effective communication. 
For example, color norms differ based on culture, knowledge, and context.  Rainbow-colored maps are typically associated with temperature variations~\cite{ma2008visitors}. If a designer applies a rainbow color map in a geospatial visualization to depict crop yields then many viewers may not properly decode the visualization.

\subsection{Context}
In communicative visualization, several works~\cite{Mukherjee2022,Schloss2019,quadri2022,Prateek2023} identified challenges in the interpretation of data-driven visualizations and how different contexts affect this interpretation.
Mukherjee et al.~\cite{Mukherjee2022} proposed the use of semantic discriminability theory, a general framework for understanding conditions determining when people can infer meaning from perceptual features.
There is a large body of linguistics research~\cite{austin1975things,searle1969speech,bach1979linguistic,sperber1986relevance,sbisa2002speech} showing how context influences the meaning of an utterance.
Sbis{\`a}~\cite{sbisa2002speech} proposes contexts are continuously shifting, but at each moment of interaction it is possible to evaluate the performed act against the context.
This literature suggests context can be classified along the following dimensions: (1) Given vs constructed context,  (2) limited vs unlimited context, and (3) context change.

\noindent\textbf{Given vs. constructed context:}
In a given context, the context is set once the event starts and is not mutable going forward.
For example, many narrative visualizations~\cite{dasu2020sea,segel2010narrative,lee2015more} or analytical systems predetermine or have a fixed context.
Whereas in a constructed context the context of an interactional event is created by its participants as the interaction proceeds.
One form of this in visualization could be highly interactive and collaborative visualizations that function off of user inputs.
These visualization evolve and change based on these interactions.
A different example of this can be seen in ~\autoref{fig:misaligned-intent} where the context of a public forum influences the designer's intent.
This begins with Jefferey Shaman creating a visualization and it is shared on a public forum.
The public became invested in whether the design is effective or not, how can it be improved, and what is the intent of this visualization.
In response to the visualization, others were created with a different intent. 
As shown in ~\autoref{fig:misaligned-intent}d, Amelia Wattenberger attempted to improve on the original, \autoref{fig:misaligned-intent}b, with some believing she did.
The constructed context in this scenario is that initially, the context of the visualization was to forecast the omicron virus for a period of time; however, as more individuals debated the effectiveness of the visualization the new visualizations produced gained a constructed context of attempting to provide an improved design and convey the original message.

\noindent\textbf{Limited vs. unlimited context:}
When is acquiring information to interpret what is occurring no longer necessary? Is the context finite or something that needs to be continuously provided?
Context, in speech act theory, has been considered a bounded resource that only includes 'what is needed' for ~\cite{kaplan1989afterthoughts} interpretation or evaluation. 
Conversely, there is an argument that context is ever-changing and that there is no end to the details one might want or need.
Searle~\cite{searle1985expression} views context as indefinitely extensible and potentially all-inclusive.
That is every speech act has meaning only against a set of background assumptions.
These assumptions are indefinite and the number of processes for the analysis and development of an idea are endless.
Other views~\cite{stalnaker1999context} find context as always extensible but delimited. 
They believe that context is needed as background information (or what the speaker believes is, or intends to treat as background information) and is delimited on every occasion by the presuppositions a speaker happens to make. 
Additionally, actions typically involve results, such as bringing about a new state, referencing the past, or substituting a new state for an older one. The objective or cognitive nature of context affects the action. 
After an action or event occurs its content is added to the participant's presuppositions, and therefore to the cognitive context. 
For example, in dashboards with linked views a user may filter on a subset of the data altering the chart.
An accompanying view may re-render to reflect this filter and reveal new insights on the subset reflecting that initial interaction.
This change is an implicit communication to the viewer that these two views are linked and the data in one is influencing the other.
The discussion of limited vs unlimited context is ongoing in speech act theory. However, the distinctions and points made, as well as points made for future works, directly apply to visualization.
For example, Mantri et al.~\cite{Prateek2023} examine how a variety of additional contexts impact the interpretation of communicative visualizations and synthesis of information that comes from consuming cumulative discoveries. 
They discuss how science and journalism often present their content as an ongoing discussion and evolution rather than a finality (i.e., discoveries that build on, contradict, contextualize, or correct prior findings).

These considerations of context clearly arise in visualization and have been defined implicitly by this classification. 
Several frameworks~\cite{lee2015more,isenberg2011collaborative,dasu2020sea,padilla2018decision} have discussed context and its influences on visualization design. 
For example, they describe external context as (1) an understanding of the target audience's needs and prior knowledge, (2) the type of device or medium the visualization will be expressed through, (3) and the physical setting.
Context's effect on inferring and interpreting visualizations has also been examined~\cite{padilla2018decision,quadri2022,Prateek2023,Schloss2019,Mukherjee2022}.
Padilla et al.~\cite{padilla2018decision} identify how after viewing the encoding, the audience mentally searches long-term memory for knowledge and context relevant to interpreting the visualization.
Furthermore, context, as it pertains to visualization, has many influences on the design and the designer's intent.
Consequently, subtle changes in intent can be reflected and seen in the visualization, as shown in~\autoref{fig:misaligned-intent}.
With VisActs, we provide a framework to facilitate studying at a granular level how intents influence the design. 

%% file: sections/04_01_visactsapplied.tex
\begin{table}[t]
\small
\centering
\begin{tabular}{p{\dimexpr 0.37\linewidth-2\tabcolsep} p{\dimexpr 0.63\linewidth-2\tabcolsep}} \hline
    \toprule
    \textbf{VisActs Terminology} & \textbf{Description} \\
    
    \arrayrulecolor{black!30}\midrule
    
    \textbf{Locutionary VisAct} & To show data. \textit{What is shared}. \\ 
    
    Data Act & 
    The curation \& selection of a dataset(s)\\

    Analytic Act &
    Expression of the data through analysis. \\

    Data Type &
    The type of data (e.g., temporal, spatial, quantitative, discrete, etc.). \\
    
    \arrayrulecolor{black!30}\midrule
    
   \textbf{Illocutionary VisAct} &
    To visualize the data. \textit{What is seen}. \\
    
    Image Act &
    The production of an image. \\

    Semiotic Act &
    The expression of data through signs \& codes. \\

    Encoding Act &
    Visual encodings mapped to data. \\

    Visualization Type &
    The type of visualization (i.e., informative, instructive, narrative, explorative, and subjective). \\

    \textbf{VisForce} &
    The designer's rationale. \\

    \arrayrulecolor{black!30}\midrule
    
    \textbf{Perlocutionary VisAct} &
    The effect the visualization has on the audience. \textit{What is understood}. \\
    
    \bottomrule
\end{tabular}
\caption{VisAct terminology. These terms and their mappings were derived from a breadth of linguistics and data visualization research.
} 
\label{tab:character_defs}
\end{table}

\section{VisActs: Application to Visualization}
To ground the value of viewing visualization as a language and applying the speech act theory we provide a set of examples.
The first example uses VisActs to asses a New York Times visualization.
These examples use VisAct at a granular level to study the intention from the perspective of two archetypes commonly observed in communicative visualization: storyteller and educator. 
As a disclaimer, we can not know for certain the original designer's intent. 
However, we can, to an extent, determine what the intent could be based on design decisions and available documentation.



\subsection{Example: Storyteller}
The storyteller is concerned with expressing data as a narrative.
Here visualization is a means to engage an audience with the data and the visualizations are carefully sequenced to illustrate causality. We apply \textit{VisActs} to study a recent narrative visualization piece, \textit{How the Virus Got Out}~\cite{wuVirusNYT}.

\begin{figure}[t]
	\includegraphics[width=\columnwidth]{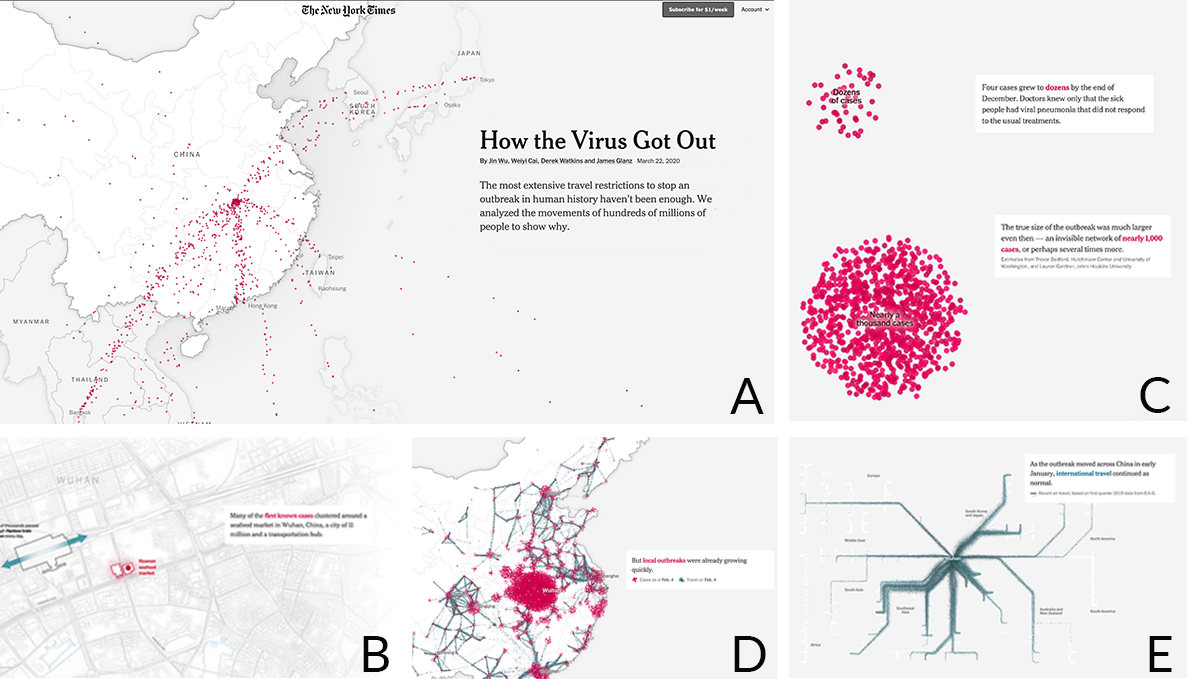}
  	\caption{\textit{"How the Virus Got Out"}~\cite{wuVirusNYT}. (a) Title page, (b)  map of Wuhan, China, (c) visualization of what was assumed to be the number of COVID cases in December compared to what it actually was, (d) overview of the virus spread, (e) scale-free representation of the world. }
  	\vspace{-3mm}
	\label{fig:nytstory}
\end{figure}

This narrative piece is composed of several visualizations, five of which are shown in ~\autoref{fig:nytstory}.
The story starts with ~\autoref{fig:nytstory}a, a visualization that promises that the authors will explain visually \textit{How the Virus Got Out} of China. 
It also asserts that the pandemic started in China.

Let us first focus on what is being shared, not inferred from the visualization. 
This is defined as the locution or locutionary act.
When contextualized as a \textit{VisAct}, the locutionary act and the locution describe what the underlying content is, namely the data. 
Here, the \textbf{locutionary VisAct} is the data and analysis used to understand the spread of the virus.
The \textit{data act} is the selection and curation of the datasets that represent people, their movements, and infection data.
As mentioned in the article~\cite{wuVirusNYT}, their data came from Baidu, two Chinese telecoms, Fred Hutchinson Cancer Research Center,  the University of Washington, and the Johns Hopkins Center
The \textit{analytic act} are the estimations and relevant methods applied to the dataset to help bring out the findings to then be visualized.
Lastly, the \textit{data types} are spatio-temporal data.

The visualizations are the illocutionary \textbf{VisAct}.
The \textbf{image act}, consists of the low-level visual elements.
The viewer would see a web page composed of color, shapes, and text (e.g. the marks and channels).
The \textbf{encoding acts} determine how marks and channels should be paired and how they are bound to the underlying data.
In this example, the data contains temporal information about people's movements, location, and estimates of the percentage of the population with COVID.
Individuals are represented as points, the position of a point connotes a geo-spatial location at an instance in time, and the color denotes whether an individual has COVID.
The meaning provided by the encoding act is supplemented by the \textbf{semiotic act}.
The semiotic act constructs the relationships between the image and other factors such as culture and society.
The grouping of shapes and text together is seen as a map of China and neighboring countries. 
This visualization uses projections and map aesthetics that most of the populace has familiarity with from map-based applications.
Therefore, the movement of points across this map is associated with transit.
In Western cultures red color has negative connotations. 
In this case, red is used to symbolize those infected with the virus.
Although the piece presents facts, because of its emphasis on temporal flow and implied causality, the \textbf{type of visualization} is a narrative. 
It is important to note that the type of visualization influences the meaning it will convey.

The \textbf{VisForce} is the intention underlying ~\autoref{fig:nytstory}a. 
One intention here is a promise to the readers \textit{to educate} how the virus spread.
A \textit{VisForce} is comprised of the seven properties described in Section 5.3.  
To understand the \textit{VisForces} at play let us first identify the set of \textbf{illocutionary points} at work. 
~\autoref{fig:nytstory}a has a commissive point and an assertive point. 
It asserts that the virus started in China and it promises to provide a justification for this assertion. 
The \textbf{degree of strength} of the promise is moderate, as we have to infer the promise. 
The mode of achievement of the \textit{VisForces} is through the sequence of steps used to build the visualization. 
It starts with the text ``China'', which slides and snaps onto a map of China.  
Then a set of red dots quickly transitions into streams of red dots flowing out of China.
The felicity conditions for these illocutionary points are the assumptions by the designers that they have the information and understanding of the base material and that the reader will benefit from the visualization.

Throughout the story, the visualizations make several assertive points about the spread of COVID, where it originated, and facts about specific days.
In ~\autoref{fig:nytstory}c, the designers present a set of points depicting the number of reported cases in December. The \textit{VisForces} consists of two assertive points and an expressive point. The illocutionary \textit{VisAct} of a small cluster of red points asserts that only a few dozen cases were known to the doctors. The second assertion was that the true number of infected was closer to a thousand.  The corresponding illocutionary \textit{VisAct} is a larger cluster.  The expressive point was the emphasis the authors placed on the difference between the two assertions.  The illocutionary \textit{VisAct} to achieve the expressive point is the animation that grows the volume of the dots.  Its degree of strength is high.

In  ~\autoref{fig:nytstory}d, volumes of varying sizes of red points are shown on a map. The designer assumes that the size of the cluster will be interpreted as the size of the infected population by the viewer.  We can argue that ~\autoref{fig:nytstory}c introduces the context necessary for this interpretation.  As we have stated earlier the \textit{VisForce} depends on the context, which can change or evolve.
The visualization in Fig~\ref{fig:nytstory}e opens with a declarative point.
There is a state change in the \textit{VisAct}. 
The overall semantic state (image, encoding, and semiotic act) of the visualization has changed.  
The visual narrative transitions from a geo-spatial visualization to a scale-free representation that mimics a subway map. 
This enables the designers to make an assertive point of Wuhan's centrality in the pandemic.

In addition to internal contexts, there are external contexts such as social conventions or norms that can passively or directly influence the \textit{VisForce}. 
For a viewer to recognize the \textit{VisForces} in this story they must be (1) aware a pandemic occurred resulting in a global lockdown, (2) familiar with reading and interpreting maps, and (3) able to understand English. 
Also, as we have seen, \textit{context change} contributes to the intended meaning. Information necessary for interpreting visual encodings is presented and then applied in different settings.
Lastly, there are some \textbf{conventions} this visualization takes advantage of.
Namely, it uses the popular scrolly-telling design and assumes those viewing the page understand the convention of scrolling to reveal new content.
The \textbf{sincerity} of the designers is seen on the final page of the story where they provide notes talking about limitations of what is presented as well as sources for the data and statements made.

To recap, we have organized the mapping of this narrative into two sections.
The first section, \textbf{locutionary acts} addresses what the designers put forth and what it is we see. 
In the second part, we focus on \textbf{illocutionary acts}. We identify (infer) the intents and examine how they are expressing their intentions. 
This is addressed by delving into the \textit{VisForces}, the illocutionary points, modes of achievement, and the context. 
In this example, the designers use several assertive illocutionary \textit{VisForces} to convey to the viewer "How the Virus Got Out".  We also identified and discussed declarative, expressive, and commissive points.

Finally, let us look at the \textbf{perlocutionary act}. 
This third and final component of \textit{VisActs} addresses the consequences of presenting the visualization to the viewer. 
The perlocutionary act captures the effect presenting the visualization had and assesses if the viewer understood the designer's intent. 
In our field, we conduct user studies and evaluations to determine this. 
To ascertain if the visualization was successful in communicating the intent of asserting the factors that led to the virus spreading across the world we would need to interview viewers.

\subsection{Example: Educator}
A common use of communicative visualization is to teach or to inform.
An educator uses visualization to simplify complex data to explain concepts. This can take place in a formal setting such as a classroom or in an informal setting like a museum.
We examine the museum exhibit DeepTree~\cite{block2012deeptree,davis2015whoa} using our framework.

Here, the \textbf{locutionary VisAct} is the phylogenetic tree of life.
The \textit{data act} is the phylogenetic dataset and the corresponding timelines.
The \textit{analytic act} could be any data formatting or association of the phylogenetic tree with the temporal information.
Lastly, the \textit{data types} are temporal and image data.

\begin{figure}[t]
	\includegraphics[width=\columnwidth]{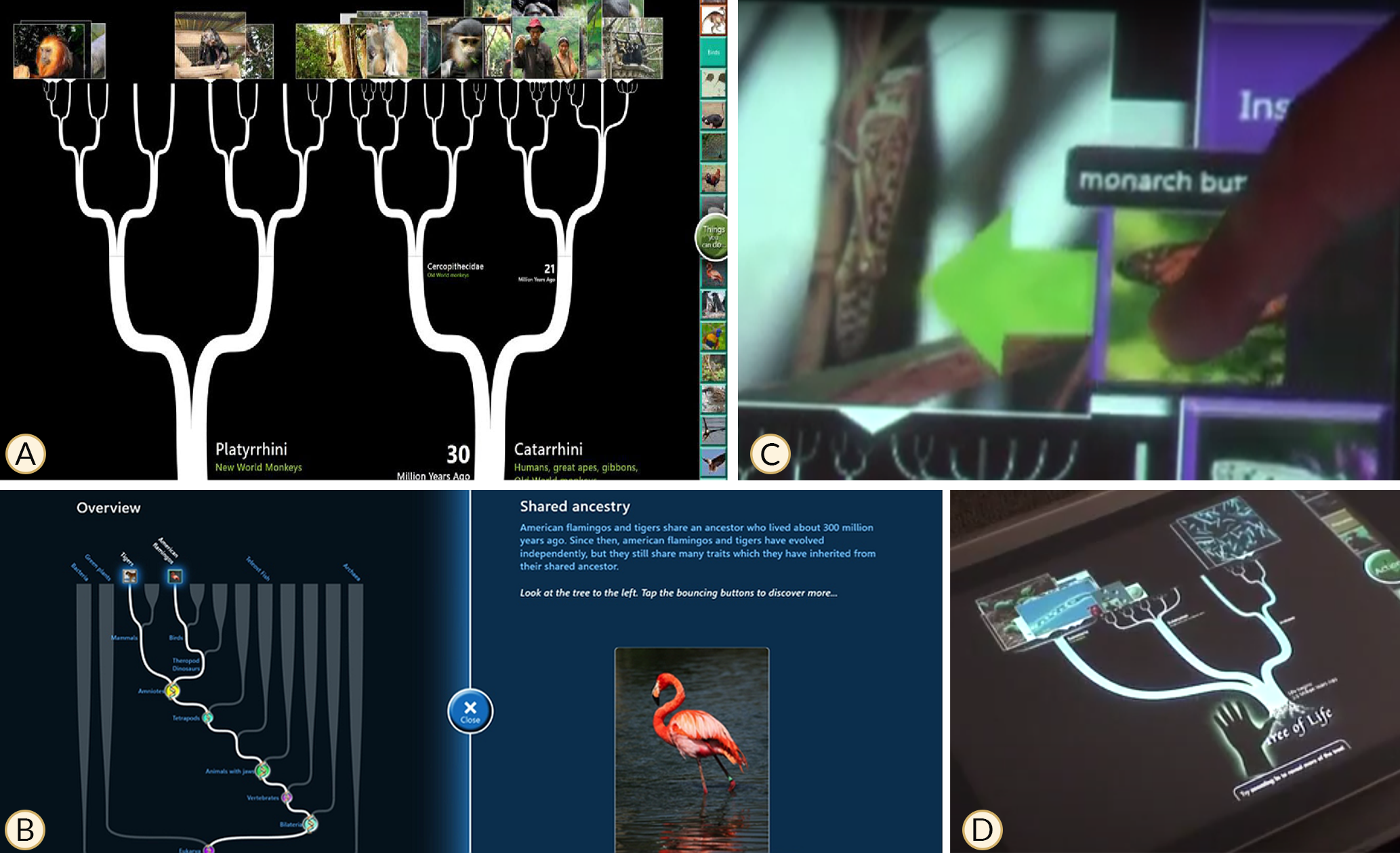}
  	\caption{ (a) DeepTree~\cite{block2012deeptree} exhibit overview (b) simplified view (c) selection interaction (d) zoom and pan interaction. (permission pending) }
  	\vspace{-3mm}
	\label{fig:eduex}
\end{figure}

DeepTree's \textbf{illocutionary act}, as seen in ~\autoref{fig:eduex}a,  has a tree-like structure supporting a set of images.
The \textbf{image act} of ~\autoref{fig:eduex}a consists of point and line marks.
DeepTree's \textbf{encoding act} maps the visual elements to the underlying data.
As a result, what is seen in ~\autoref{fig:eduex}a is a phylogenetic tree dataset that consists of images and text.
The designers create a tree visualization where leaves are images of species. 
The visualization is composed of a line mark and takes advantage of size and position channels to convey the dataset.
Images on the tree depict species along with a label with their common name.
The \textbf{semiotic act} provides a tree metaphor, where the trunk symbolizes the universal common ancestor and the branches the children. The "branches" of this tree represent splits in species and portray the phylogenetic classification.
This \textbf{type of visualization} is primarily explorative and is supplemented with some informative elements.

As with the prior example, to understand the illocution, we identify the \textit{VisForces}.
This visualization makes use of four of the five illocutionary points as described in Table~\ref{tab:speech_vis}.  It has assertive, directive, commissive, and declarative points.
We will focus on the directive and commissive points.

\autoref{fig:eduex}a is the entry point for this visualization.  The \textit{VisForces} here include commisive and directive points. The \textit{VisAct} promises to inform the visitor about the Tree of Life.   The mode of achievement is a collection of images of different species and animations indicating relationships among them.  The directive point is to get the viewers to drag their hands across the screen. 
The mode of achievement is an animation of a hand dragging downward to instruct viewers how to engage with the application, ~\autoref{fig:eduex}d.
This visualization has several other directive points that are achieved through different modes such as tapping a button, downward dragging to traverse down the tree, pushing upward to move up the tree, flicking to quickly move through the tree, single-touch pan,  multi-touch zoom, pushing to select, and dragging an element onto a destination.
Each directive point is achieved by using techniques such as highlighting, animating, or annotating visual elements to cue the viewer to interact.
The degree of strength for a directive point depends on the visual emphasis placed on that technique.

External factors and conventions also influence the directive points.
The museum setting and use of an interactive touch-screen table to display the visualization add to the \textit{VisForce}.

The side panel in ~\autoref{fig:eduex}a has a commissive point. The promise here is to inform the viewer of the location of the tree of each species portrayed in the side panel.  In DeepTree when a user selects an image of a species from the side-panel, ~\autoref{fig:eduex}a, the image jumps to its position in the tree. 
If an image is pressed a graphical element, an arrow, appears showing the viewer where to slide (~\autoref{fig:eduex}c). 
The directive point here is to get the viewer to slide the image.
This directive point is weaker than the earlier directive point for getting a viewer to drag their hand onto the table.

Let us next examine the \textbf{propositional content condition} for directive points.  These are the designers' beliefs that the viewer will perform an action they request.
In DeepTree, the designers believe that by animating an element to grow and shrink, adding a highlight around it, and having a text annotation above it saying ``learn more'' the viewer will tap on it.
The \textbf{preparatory conditions} for all directive points assume that the viewer is able to perform the suggested actions.
The \textbf{sincerity condition} of these directive points is that the designer wants the viewer to perform the actions.
The degree of strength for the sincerity condition is the importance of these actions to the designer.
In DeepTree it is very important for the designers that the viewers pan and navigate the tree. This action is crucial and is evidenced by the visual emphasis placed on this \textit{VisAct}.

The viewer can press a button that takes them to a new simplified view of the tree.
The illocutionary \textit{VisAct} here is a tree of life contextualized to the selected species, ~\autoref{fig:eduex}b.
The \textit{VisForces} here have a declarative and commissive point.
The declarative point is the transition from the earlier state, ~\autoref{fig:eduex}a,  to a simplified visualization, ~\autoref{fig:eduex}b.
This declarative point's mode of achievement is an animated transition.

The commissive point the simplified view makes is that it promises the viewer that the designer will return them back to the original state, ~\autoref{fig:eduex}a.
The \textbf{mode of achievement} for this commissive point is a button. 
The \textbf{propositional content condition} for this commissive point is that the designer will fulfill the commitment.
That is, upon a viewer pressing the ``X'' button, seen in ~\autoref{fig:eduex}b, the simplified view will disappear and "close" and they will be returned back to the overview, ~\autoref{fig:eduex}a.
The \textbf{preparatory conditions} for the commissive point is that the designer is able to complete this promise within their design.
The \textbf{sincerity condition} of this commissive point is that designer, and therefore the visualization intends to satisfy the promise.

Briefly, there are many assertive made in both the simplified and overview visualizations.
These assertive points state facts about species and their ancestry. 
The modes of achievement the designers selected to express their assertive points include dialog/pop-up boxes, annotations, and color to highlight relationships between species. 

So far, we have gone over some of the \textit{VisForces} present in this exhibit to illustrate how to use our framework and the structure it provides.
Namely, we showed that there are assertive, declarative, commissive, and directive points and thus those respective \textit{VisForces}.
We walked through the properties of some of these forces and gave examples (i.e. we described eight directive \textit{VisForces}, degree of strength, conditions, and mode of achievement).
However, we also have to account for how social conventions and external contexts influence the visualization design and its overall meaning.

DeepTree relies on external contexts and conventions present in an informal learning environment;
specifically a museum and the considerations and conventions~\cite{dasu2020sea} that it comes with. 
Furthermore, DeepTree relies on its viewers to have familiarity with touch-based devices~\cite{block2012deeptree} (e.g., iPads and iPhones).

Lastly, the perlocutionary \textit{VisAct} addresses the reaction the viewers had upon seeing the visualization.
It can be used to determine if the designer was successful in conveying their intended meaning.
The designers of DeepTree documented their evaluation~\cite{block2012deeptree} and from it we can see that their  directive and a comissive \textit{VisForces} were understood by the viewer.
For example, both the commissive force of a promise to the viewer that by tapping the find button something will occur in the future and the directive force of a request to the user to tap and drag an image off the image reel were successful.
Viewers would tap on the button to find a species, signifying the viewer believes a promise will be fulfilled.
Additionally, they were then presented with a slot to place an image.
They inferred the directive point and dragged an image onto the slot.
After doing so, the promise made by the designer is fulfilled as the visualization ``flies'' through the tree via animation to where the species in the image is located.
This \textit{VisAct} had  ``emotional'' perlocutionary response in the viewers, where the designers documented responses such as ``wow, this is big'' or ``woah''. 

%% file: sections/05_discussion.tex
\section{Discussion}
With VisActs, we present a conceptual framework for inferring designer intent.
This framework is not a finality but rather a foundation to be iterated and expanded upon.
There is a growing need to infer and assess designer intent, as well as grow our understanding of which types of design decisions illustrate an intent (i.e., negative emotions can be conveyed via a variety of design choices~\cite{NegativeEmo2022}).
With this manuscript, we provide a translation and contextualization of frameworks from linguistics to communicative visualization and offer a framework for inferring intent in interactive data-driven visualizations.
In this section, we discuss the prospective values and directions VisActs can lead to.


\begin{figure}[t]
	\includegraphics[width=\columnwidth]{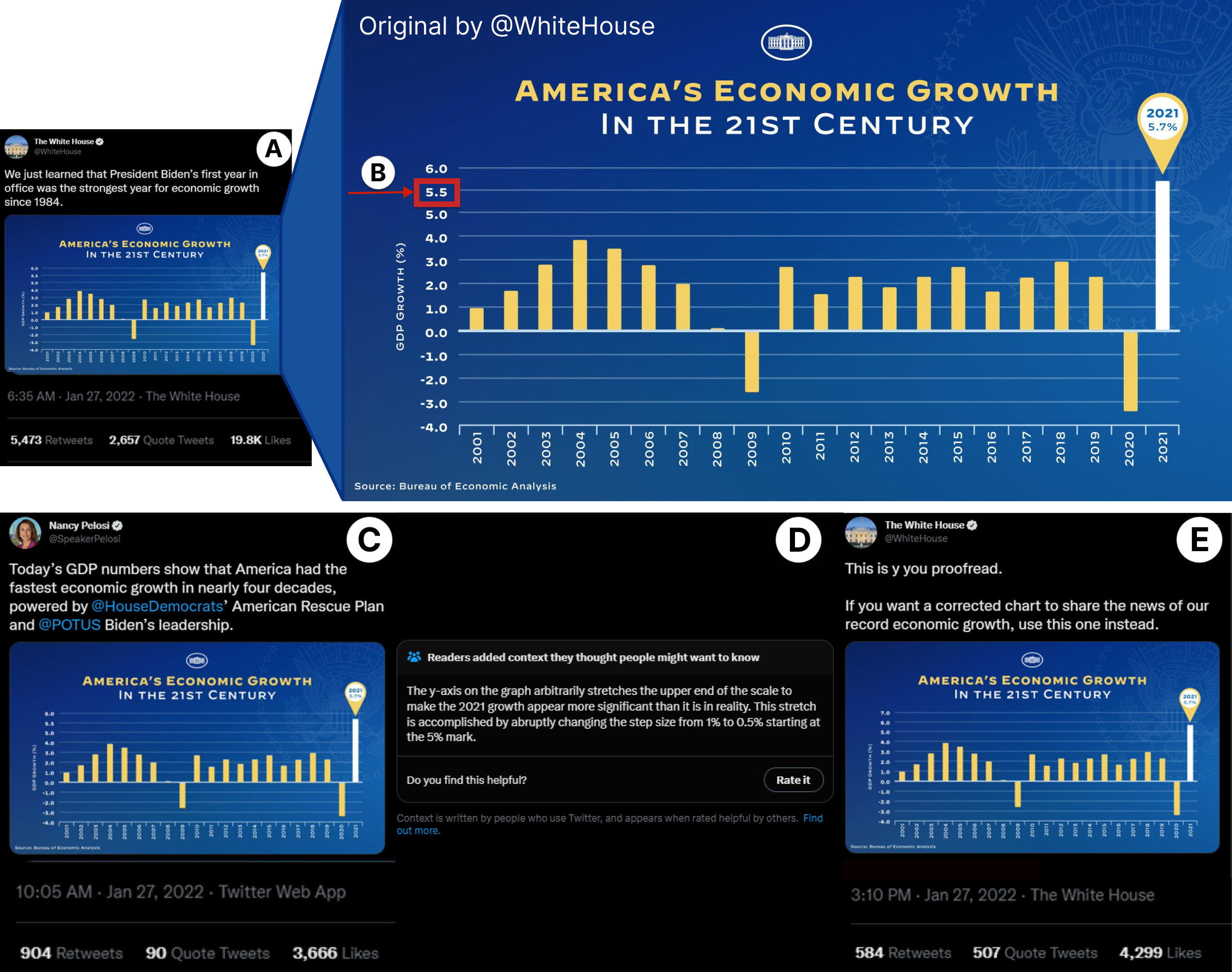}
  	\caption{ (a) Tweet by the @WhiteHouse account conveying the economic growth of the current administration~\cite{whitehouse2021}. (b) The y--axis label increments by a half-step rather than a whole point as it was previously. (c) Nancy Pelosi shares this figure with the error to her following~\cite{pelosi2021} (8.1 million followers). (d) The Twitter community flags the discrepancy and many replies and threads are made questioning the integrity of the data. (e) @WhiteHouse updates the figure stating it was a \textit{proofreading} issue. }
  	\vspace{-3mm}
	\label{fig:mistakeOrIntent}
\end{figure}

\noindent\textbf{Biases Assessment:} One use of VisActs is to tease out design decisions that could reflect the designer's biases. For example, in \autoref{fig:mistakeOrIntent}a we see a bar chart communicating \textit{``America's economic growth in the 21st century''}.
This chart at a glance shows the tallest bar to be for the year 2021. 
Upon closer review, however, \autoref{fig:mistakeOrIntent}b, we see that the y--axis has a mistake.
This mistake extends the 2021 bar to be slightly more exaggerated in comparison to the other years.
This mistake was shared for several hours on Twitter before a correction was made.
In that span, the community responded with many complaints and comments over the chart.
There were claims that it was intentionally added as a means to persuade that the 2021 administration is very effective.
The public would attempt to evidence these claims by suggesting the mistake only occurred at a point on the y-axis that would only affect the 2021 bar and nothing else.
Namely, the public assessed the encoding design to infer a plausible designer intent. 

It is impossible to say with absolute certainty whether this case was a genuine mistake or an intentional design choice.
However, there is a need to provide structures to assess and better debate the biases in charts that circulate in the wild.
As these affect how society trusts data and science.
One possible extension of VisActs is to apply the framework to a corpus of data visualizations and create associations between design and possible intents.
Through developing richer classification models, we could develop faster or more accurate linters that could identify these discrepancies and provide community notes, \autoref{fig:mistakeOrIntent}d.

\noindent \textbf{ML4Vis application.} 
Bako et al.~\cite{Bako2022} call attention to how visualization examples can be used as inputs to author visualizations as a direction for future work.
To help achieve this goal, we need to really understand and expand on which aspects of these examples correlate to what the designer intends to do.
This could be task-based as well, however, it is possible that task-based may not be granular enough to effectively capture the needs of designers and the specific design elements they may be interested in.
Expanding on classification models VisActs has an application in the ML4Vis space. 
Principles and frameworks from speech act and discourse theory have been studied and leveraged in NLP. Similarly, VisActs can be utilized in future works to automatically generate visualizations and their design based on designer inputs.  
By offering a framework to infer the designer's intentions based on design rationale and assess which features could contribute to particular intentions we can better machine learning train models on data visualization and build associations between visual features and particular intentions.

As with Midjourney and DallE prompts, VisActs can be a stepping stone to developing applications that allow users to provide their data to visualize and enter prompts to tailor the visualization to their needs. 
In order to achieve such automation we need to develop rich associations between designer intents and corresponding design choices.
Through VisActs, it is possible to develop these associations. This framework can be applied at a granular level, as seen in the examples for Storyteller and Educator.

%% file: sections/06_conclusion.tex
\section{Conclusion}
This work takes the view that visualization is a language and can therefore benefit from applying frameworks and theories from linguistics to systematically understand and analyze how we communicate through data visualization.
We provide a translation of a sub-field of linguistics and offer our framework VisActs.
We then use examples applying our framework to illustrate its potential application to our field.
This translation affords us a means to deconstruct the language of visualization, identify low-level communicative components, and learn how these components individually and collectively accomplish the communicative goal of the visualization. 
Our detailed examples demonstrate how these concepts can be used to examine designer intents and describe the forces at play. 

This is an initial mapping of the two spaces and future work can tighten this association and build upon its structure.
We believe that our work gives credence to the relevance of linguistics frameworks for the study of visualization and supports continued efforts in translating other frameworks and theories into our domain.
We hope our work enables the future integration of theories and frameworks from linguistics into visualization and grows our framework for studying visualization design intent.
VisAct provides a standard way, a language, for anybody to examine/describe a visualization, its interaction design, and the designer's intent down to the granular level.